\documentclass[a4paper,20pt]{article}
    \usepackage[T1]{fontenc}
    \usepackage{graphicx}
    \usepackage{epsfig}
    \usepackage{amsmath}
    \usepackage{amsfonts}
    \usepackage{hhline}%
    \usepackage{textcomp}
    \renewcommand{\abstract}{}
    \newcommand\sun{\hbox{$\odot$}}
\begin{document}
\makeatletter
\renewcommand{\@oddhead}{\textit{YSC'16 Proceedings of Contributed Papers} \hfil \textit{R. I. Uklein}}
\renewcommand{\@evenfoot}{\hfil \thepage \hfil}
\renewcommand{\@oddfoot}{\hfil \thepage \hfil}
\fontsize{11}{11} \selectfont

\title{3D-structure of the Canes Venatici I Cloud}
\author{\textsl{R.~I.~Uklein}}
\date{}
\maketitle
\begin{center} {\small Special Astrophysical Observatory RAS, 369167 Nizhny Arkhyz, Karachay-Cherkessia, Russia\\
                       uklein@sao.ru}
\end{center}

\begin{abstract}
We present the improved distance moduli of 30 galaxies in the Canes Venatici I Cloud using advanced Tip of Red Giant Branch (TRGB) method \cite{makarov06}. The method was determined for accurate estimation of the distances even if TRGB situated near photometric limit. The data were taken from the Archive of the Hubble Space Telescope (HST). Based on ACS and WFPC2 images of the HST we construct the color-magnitude diagrams of the resolved stellar population of the galaxies using Dolphot and HSTPhot packages.

New refined method of the distance determination allows us to clarify the 3D structure of the Canes Venatici I Cloud. It consists of the central group of galaxies around M94 and the outskirt which is situated in     gravitational field of the ``core''. The mass and mass-to-light ratio of the CVn  have been estimated.
\end{abstract}

\section*{Introduction}
\indent \indent Determination of distances is important task in astrophysics, especially in extragalactic astrophysics. For the study of structure, kinematics and dynamics of galaxy groups and clusters we must know accurate distances to galaxies. Our object is a scattered cloud of galaxies in Canes Venatici constellation, CVn I Cloud. The brightest galaxy of the cloud is M94 (other name of this complex is M94 group). Neighboring galaxy groups (Local Group, M81, Cen A and others) are compact dense groups with strong concentration to central massive galaxy. Unlike those groups, the CVn I Cloud is scattered, sparse grouping in the Local Volume. In our work we derived the distances to CVn I Cloud galaxies from the luminosity of the tip of the RGB stars.

The CVn I Cloud is clearly seen in redshift space. Histogram in Figure~1 shows a rather isolated peak at $V_{LG} = 200-350$~km/s, which caused by galaxies in the CVn I Cloud. For our sample we selected the galaxies with radial velocities less than 400~km/s.

\section*{Method of the distance determination}
\indent \indent For distance determination we use advanced Tip of the Red Giant Branch method \cite{makarov06}. This method is based on a maximum likelihood algorithm of locating the TRGB position and is optimized by introducing reliable photometric errors and completeness function determined from artificial star experiments. Advanced TRGB-method is especially useful in the cases when the TRGB approaches the photometric limit or is poorly populated.

The source of images is snapshot programs of the Hubble Space Telescope (HST). The observed photometry is extracted from HST WFPC2 images with HSTphot package or from HST ACS images with the DOLHPOT package \cite{dolphin00}. Obtained Color-Magnitude Diagrams (CMDs) are analyze with the TRGBTOOL program developed by Makarov for realization of  advanced TRGB-method. 

The results of distance determination are summarized in Table 1. This table contains the following columns: 
(1) galaxy name; 
(2) equatorial coordinates of the galaxy center (J2000);
(3) radial velocity with respect to the LG centroid;
(4) distance to galaxy in Mpc;
(5) positive and negative uncertainty of distance in Mpc;
(6) blue absolute magnitude of a galaxy with the given distance in magnitudes;
(7) morphological type in de Vaucouleurs notation.
Typical accuracy of the most of our distances is 5-6\%. Such result places confident CMDs with considerable stellar population in the TRGB region with TRGB situated at more then 1$^m$ above the photometric limit.
For some CMDs distance errors may be very large, up to 25\%.

\begin{table}[t]
\centering
\caption{Properties of CVn I galaxies.}
\centering
\begin{tabular}[h]{lcccccr}
\hline
Name & RA (J2000) Dec & $V_{LG}$ & $D$, Mpc & Error, Mpc & $M_B$, mag & Type \\
\hline
UGC6817 & 115053.0$+$385250 & 248 & 2.67 & $+$0.15/$-$0.14 & $-$13.52 & 10 \\
UGC6541 & 113329.1$+$491417 & 304 & 4.29 & $+$0.29/$-$0.27 & $-$13.71 & 10 \\
NGC3738 & 113548.6$+$543122 & 305 & 3.66 &  & $-$16.61 & 10 \\
NGC3741 & 113606.4$+$451707 & 264 & 3.14 & $+$0.19/$-$0.18 & $-$13.13 & 10 \\
NGC4068 & 120402.4$+$523519 & 290 & 4.27 & $+$0.21/$-$0.20 & $-$15.07 & 10 \\
NGC4163 & 121208.9$+$361010 & 164 & 2.92 & $+$0.15/$-$0.14 & $-$13.81 & 9 \\
UGCA276 & 121457.9$+$361308 & 283 & 2.96 & $+$0.18/$-$0.16 & $-$11.67 & $-$1 \\
NGC4214 & 121538.9$+$361939 & 295 & 3.26 & $+$0.19/$-$0.17 & $-$17.19 & 9 \\
UGC7298 & 121628.6$+$521338 & 255 & 4.32 &  & $-$12.27 & 10 \\
NGC4244 & 121729.9$+$374827 & 257 & 4.3 & $+$0.29/$-$0.25 & $-$18.6 & 6 \\
UGC7559 & 122705.1$+$370833 & 225 & 4.71 & $+$0.26/$-$0.25 & $-$14.38 & 10 \\
UGC7577 & 122741.8$+$432938 & 237 & 2.62 & $+$0.13/$-$0.12 & $-$14.16 & 10 \\
NGC4449 & 122811.2$+$440540 & 246 & 4.31 & $+$0.24/$-$0.23 & $-$18.27 & 9 \\
UGC7605 & 122839.0$+$354305 & 312 & 4.62 &  & $-$13.53 & 10 \\
IC3687  & 124215.1$+$383007 & 369 & 4.15 & $+$0.23/$-$0.21 & $-$14.64 & 10 \\
KK166   & 124913.3$+$353645 &     & 4.48 &  & $-$10.82 & $-$3 \\
M94 & 125053.5$+$410710 & 353 & 4.82 &  & $-$19.83 & 2 \\
IC4182  & 130549.3$+$373621 & 356 & 4.43 & $+$0.04/$-$0.03 & $-$16.4 & 9 \\
UGC8215 & 130803.6$+$464941 & 297 & 4.36 & $+$0.24/$-$0.22 & $-$12.3 & 10 \\
UGC8308 & 131322.8$+$461911 & 243 & 4.17 & $+$0.52/$-$0.42 & $-$12.7 & 10 \\
UGC8320 & 131428.6$+$455510 & 273 & 4.22 & $+$0.24/$-$0.23 & $-$15.28 & 10 \\
NGC5204 & 132936.4$+$582504 & 341 & 3.32 &  & $-$16.75 & 9 \\
UGC8508 & 133044.4$+$545436 & 186 & 2.68 & $+$0.14/$-$0.14 & $-$12.98 & 10 \\
UGC8638 & 133919.4$+$244633 & 273 & 4.24 & $+$0.21/$-$0.20 & $-$13.77 & 9 \\
KK230   & 140710.7$+$350337 & 126 & 1.98 & $+$0.17/$-$0.25 & $-$9.57 & 10 \\
UGC8331 & 131530.7$+$472947 & 345 & 4.39 &  & $-$13.71 & 10 \\
UGC8651 & 133953.8$+$404421 & 272 & 3.07 &  & $-$12.97 & 10 \\
UGC8760 & 135051.1$+$380116 & 257 & 3.22 &  & $-$13.13 & 10 \\
UGC8833 & 135448.7$+$355015 & 285 & 3.06 &  & $-$12.37 & 10 \\
UGC9128 & 141556.5$+$230319 & 172 & 2.24 &  & $-$12.51 & 10 \\
\hline
\end{tabular}

\end{table}

\section*{Results}
\indent \indent Based on the obtained results we build the 3D distribution of our galaxies in Supergalactic coordinates (Figure~2). It is clearly seen that CVn I Cloud is an elongated complex of galaxies.

New refined distances allow us to clarify the structure of the complex. The cloud is divided into the ``core'' and the ``outskirt''. The core contains the brightest galaxies: M94, NGC 4449, and NGC 4244. It is worth to note, that those galaxies have luminosities by an order of magnitude less than luminosities of the brightest galaxies in the nearby groups -- Milky Way, M31, M81 and other central galaxies. Periphery galaxies, predominantly of irregular type, lie near zero-velocity surface of the core gravitational field.

Figure 3 presents the Hubble diagram for galaxies in the CVn Cloud. Black circles corresponds to our distance estimations and open circles show the measurements from \cite{karachentsev03}. They used TRGB method for 18 galaxies with precision  $\sim 12\%$ and compiled 54 estimations from previous  works. Most of the distances above 5~Mpc were measured by the brightest stars method which has precision less than~20\%.

Renewed Hubble diagram allows us to make several conclusions. As it is shown in Figure~3, galaxies of the core are groupping to narrow band at distance 4.3~Mpc. This feature is typical for virialized groups of galaxies.
Probably, the core is virialized system too. But we are sure that the cloud itself does not reach virialized state. Regression line shows pure Hubble law. The value of the Hubble constant is equal to 72~km~s$^{-1}$~Mpc$^{-1}$. The median velocity of the core is about 300~km/s.

The new Hubble diagram makes it possible to estimate the radius of zero-velocity surface. This surface divides space into inner zone of gravitational influence of galaxies and outer zone, where Hubble flow is dominated. The zero-velocity radius for our cloud is about 1~Mpc. It is interesting but the radii of zero-velocity surface of the nearby groups have the same value about 1~Mpc.

We estimated the total mass and mass-to-luminosity ratio for the CVn I Cloud core. For this we suppose, that the mass is concentrated in the core galaxies, because they are the most massive ones. We obtained the total mass of about $3\cdot 10^{12}$ solar masses using different methods from \cite{heisler85}. The results are presented in Table 2.

\begin{table}[t]
\centering
\caption{Total mass estimation of CVn I Cloud}
\begin{tabular}[h]{lc}
\hline
Method & Total Mass, $10^{12} M_{\sun}$ \\
\hline
Zero-Velocity Surface & 2.1 \\
Virial Theorem & 4.9 \\
Projected Mass & 2.7 \\
Average Mass & 3.0 \\
Median Mass &  4.6 \\
\hline
\end{tabular}
\end{table}

The total blue luminosity of the core is equal to about $2\cdot 10^{10} M_{\sun}$. It gives the mass-to-luminosity ratio for the CVn of 100$-$200~$M_{\sun} /L_{\sun}$. It is a rather big value, that indicates the presence of significant amount of the dark matter.

\section*{Conclusions}
\indent \indent The homogeneous distances to the galaxies in the CVn I Cloud were determined using improved TRGB method. Most of the measurements have 5-6\% error. Available images of the Hubble Data Archive allow us to measure distances up to $5-6$~Mpc. We need special observations with long exposure on new WFPC3 camera to reach RGB stars for more distant galaxies.

New data allow us to distinguish the core and foreground outskirt of the CVn Cloud. Unfortunately, we cannot say anything about rear of the cloud because that galaxies lie beyond the limit of the accurate distance determination.

Though this cloud does not contain giant galaxies and almost all galaxies are of dwarf irregulars the dynamical mass of the CVn I Cloud is nearly the same as masses of groups like Local Group with massive central galaxies. Most likely the CVn I cloud is semi-virialized group of galaxies formed in the gravitational field of the dark matter with center at distance $4-5$~Mpc in the Canes Venatici constellation. 

\section*{Acknowledgments}
This work was supported by RFBF grants 08-02-00627.

\begin{figure}[p]
\centering
\epsfig{figure = 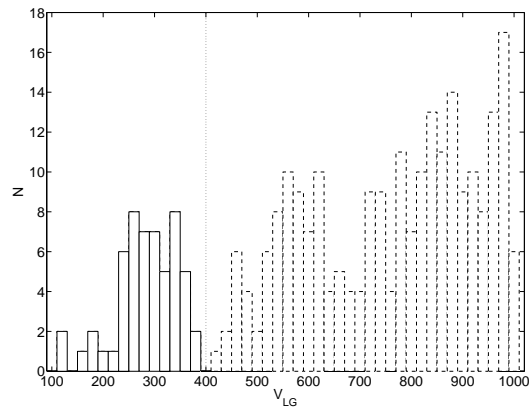,width = .45\linewidth}
\caption{Radial velocity histogram of galaxies in the CVn region.}
\end{figure}
\begin{figure}
\centering
\epsfig{figure = 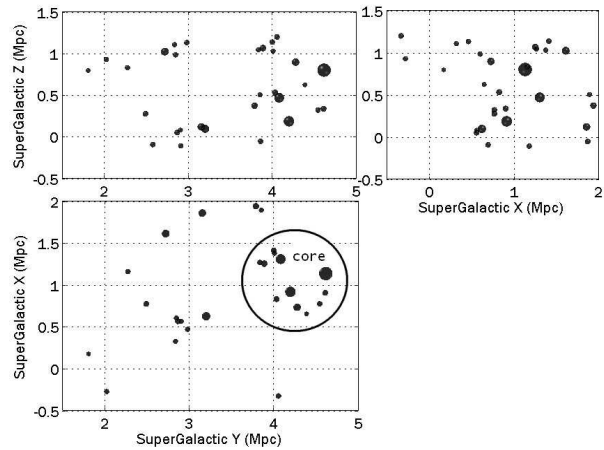,width = .45\linewidth}
\caption{Distibution in Supergalactic coordinates. Size of the circle corresponds to luminosity of galaxy. Galaxies of the core are encircled.}
\end{figure}
\begin{figure}
\centering
\epsfig{figure = 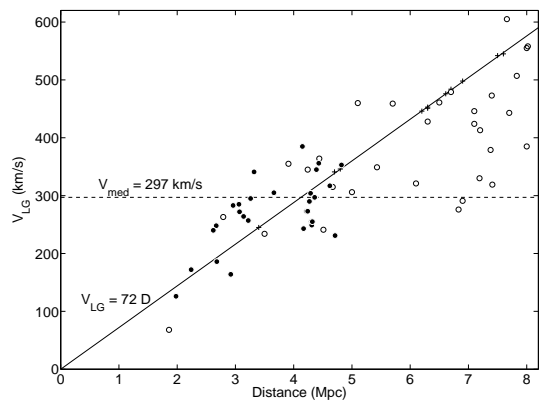,width = .45\linewidth}
\caption{Hubble diagram for galaxies in the CVn I region. The galaxies with new distances are indicated by filled black circles, with old distances -- by unfilled circles, without distances -- by  crosses (Hubble flow distances). The regression line corresponds to the Hubble constant $H_0 = 72$~km~s$^{-1}$~Mpc$^{-1}$}
\end{figure}

\end{document}